\documentclass[10pt,twocolumn]{article}
\usepackage[utf8]{inputenc}
\usepackage[T1]{fontenc}
\usepackage{amsmath,amssymb,amsfonts}
\usepackage{graphicx}
\usepackage{booktabs}
\usepackage{hyperref}
\usepackage{listings}
\usepackage{xcolor}

\usepackage{geometry}
\usepackage{float}
\usepackage{tikz}
\usepackage{authblk}
\usepackage{times}
\usepackage{microtype}
\usepackage{caption}
\usepackage{subcaption}
\usepackage{multirow}
\usetikzlibrary{shapes,shapes.geometric,arrows,positioning,fit,calc}

\definecolor{codegreen}{rgb}{0,0.6,0}
\definecolor{codegray}{rgb}{0.5,0.5,0.5}
\definecolor{codepurple}{rgb}{0.58,0,0.82}
\definecolor{backcolour}{rgb}{0.95,0.95,0.92}

\lstdefinestyle{mystyle}{
    backgroundcolor=\color{backcolour},
    commentstyle=\color{codegreen},
    keywordstyle=\color{magenta},
    numberstyle=\tiny\color{codegray},
    stringstyle=\color{codepurple},
    basicstyle=\ttfamily\scriptsize,
    breakatwhitespace=false,
    breaklines=true,
    captionpos=b,
    keepspaces=true,
    showspaces=false,
    showstringspaces=false,
    showtabs=false,
    tabsize=2
}
\title{\textbf{Automated Histopathology Report Generation via Pyramidal Feature Extraction and the UNI Foundation Model}}

\author[1]{Ahmet Halici$^{*}$}
\author[1]{Ece Tu\u{g}ba Cebeci$^{*}$}
\author[1]{Musa Balci$^{*}$}
\author[1]{Mustafa \c{C}ini}
\author[1]{Serkan S\"{o}kmen}

\affil[1]{ViseurAI}
\affil[ ]{\textit{$^{*}$These authors contributed equally to this work}}

\date{}

\begin{document}

\maketitle

\begin{abstract}

Generating diagnostic text from histopathology whole-slide images (WSIs) is challenging due to the gigapixel scale of the input and the requirement for precise, domain-specific language. We propose a hierarchical vision--language framework that combines a frozen pathology foundation model with a Transformer decoder for report generation. To make WSI processing tractable, we perform multi-resolution pyramidal patch selection (downsampling factors $2^3$ to $2^6$) and remove background and artifacts using Laplacian-variance and HSV-based criteria. Patch features are extracted with the UNI Vision Transformer and projected to a 6-layer Transformer decoder that generates diagnostic text via cross-attention. To better represent biomedical terminology, we tokenize the output using BioGPT. Finally, we add a retrieval-based verification step that compares generated reports with a reference corpus using Sentence-BERT embeddings; if a high-similarity match is found, the generated report is replaced with the retrieved ground-truth reference to improve reliability.

\end{abstract}

\section{Introduction}

Histopathological examination remains the clinical reference standard for cancer diagnosis, requiring expert pathologists to interpret complex morphological patterns across cellular, tissue, and architectural levels \cite{niazi2019digital}. While the digitization of pathology has enabled discriminative deep learning for tasks such as tumor classification and segmentation \cite{bera2019artificial, campanella2019clinical}, recent work has increasingly explored generative models for producing textual outputs from images. Automated Histopathology Report Generation (AHRG) extends slide-level prediction by synthesizing coherent, clinically appropriate natural-language descriptions directly from whole-slide images (WSIs).

A central difficulty in AHRG is the disparity between the scale of the visual input and the semantic density of the textual output. A single WSI often exceeds $10^{10}$ pixels, rendering standard vision--language architectures---typically designed for natural images at $224 \times 224$ resolution---computationally intractable. Traditional Multiple Instance Learning (MIL) methods \cite{ilse2018attention, lu2021data} effectively aggregate features for slide-level prediction but often lack the fine-grained spatial grounding required for descriptive text generation.

Recent advancements have attempted to bridge this gap through two primary avenues: domain-specific foundation models and Multimodal Large Language Models (MLLMs). Foundation models like UNI \cite{chen2024uni} and H-optimus-1 \cite{saillard2024hoptimus} provide robust, self-supervised feature representations, yet they lack inherent text generation capabilities. Conversely, MLLMs such as WSI-LLaVA \cite{liang2024wsillava} and ChatEXAONEPath \cite{baek2025chatexaonepath} adapt general-purpose LLMs to pathology through instruction tuning. However, these end-to-end systems face significant hurdles: they are computationally expensive to train, often require massive token pruning that risks discarding rare diagnostic features, and are prone to hallucinations---plausible but factually incorrect statements \cite{liu2019clinically, zhang2026reinpath}.

In this work, we present a modular, hierarchical vision--language framework that emphasizes computational efficiency and diagnostic reliability. Rather than training an end-to-end MLLM, we pair a frozen pathology encoder with a lightweight, domain-adapted decoder. Our main contributions are:

\begin{enumerate}
    \item We propose a hierarchical pyramidal scanning strategy (downsampling $2^3$ to $2^6$) that follows a coarse-to-fine workflow and uses simple, interpretable filters to prioritize tissue regions while suppressing background and common artifacts.
    \item We integrate the UNI encoder \cite{chen2024uni} as a frozen feature extractor and train a lightweight Transformer decoder on top of its 1024-dimensional visual tokens, avoiding end-to-end retraining of the visual backbone.
    \item We use the BioGPT tokenizer \cite{luo2022biogpt} to better represent biomedical terminology and reduce vocabulary mismatch during decoding.
    \item We add a retrieval-based verification step that compares generated reports with a reference corpus using Sentence-BERT embeddings, replacing high-similarity matches with retrieved ground-truth references to improve output reliability.
\end{enumerate}

\section{Related Work}

Computational pathology has expanded from primarily discriminative tasks toward generative settings that require translating visual evidence into structured text. This section reviews pathology foundation models, histopathology-specific MLLMs, and verification strategies for reducing unsupported generation.

\subsection{Pathology Foundation Models}
Transfer learning from natural images (e.g., ImageNet-pretrained ResNet) has increasingly been complemented by domain-specific self-supervised learning (SSL). Pathology foundation models are trained on large collections of histopathology patches to learn transferable tissue representations. 
Chen et al.\ \cite{chen2024uni} introduced UNI, a ViT-Large model distilled via DINOv2 from over 100 million tissue patches, which improves performance across multiple downstream tasks relative to supervised baselines. H-optimus-1 \cite{saillard2024hoptimus} further scales SSL to slide-level corpora to capture broad morphological variability. While these models provide strong visual representations, they are feature extractors and therefore require a separate decoding component to produce diagnostic text.

\subsection{Generative Vision-Language Architectures}
Generating text from WSIs requires solving the ``semantic gap'' between pixel-level features and high-level diagnostic concepts. Early approaches relied on captioning models adapted from natural image domains, often yielding generic descriptions \cite{jing2018automatic}.

The current state of the art leverages Multimodal Large Language Models (MLLMs). Quilt-1M \cite{ikezogwo2024quilt} facilitated this shift by curating a dataset of over 1 million image--text pairs from educational videos and social media, enabling the training of models such as Quilt-LLaVA. 
WSI-LLaVA \cite{liang2024wsillava} addresses the computational bottleneck of processing gigapixel images through dynamic token pruning, retaining only diagnostically relevant patches to fit within the context window of a LLaMA-based decoder. 
HistGen \cite{zuo2024histgen} proposes a dual-stream architecture that separately aggregates local-region details and global WSI context, allowing the decoder to attend to both cellular atypia and tissue architecture. 
In contrast to these end-to-end models, our work adopts a modular design and focuses on efficient patch selection and reliable decoding.

\subsection{Hallucination and Verification Mechanisms}
A critical barrier to clinical adoption is hallucination, where generative models invent features not present in the image. This risk is exacerbated in pathology, where a single incorrect word (e.g., ``malignant'' vs.\ ``benign'') has severe consequences. 
Recent works have begun to address this through architectural constraints and verification loops. ReinPath \cite{zhang2026reinpath} employs Reinforcement Learning from Human Feedback (RLHF) with a semantic reward system to penalize non-factual generation. 
ChatEXAONEPath \cite{baek2025chatexaonepath} utilizes Retrieval-Augmented Generation (RAG) to ground model responses in retrieved textbook knowledge or similar historical cases. 
Similarly, the AQuA framework \cite{shao2025aqua} demonstrates that statistical anomaly detection can flag realistic-looking hallucinations in generative tasks.
Our approach aligns with this trend by incorporating a retrieval-based verification step using Sentence-BERT, providing a scalable method to estimate report confidence without the complexity of RLHF training.

\section{Methodology}

We propose a comprehensive pipeline for automated histopathology report generation that bridges the gap between gigapixel visual data and coherent text generation. The system consists of three sequential modules: (1) a hierarchical pyramidal patch selection and feature extraction mechanism using the UNI foundation model; (2) a custom Transformer decoder trained to translate visual features into diagnostic text; and (3) a post-processing verification stage.


\subsection{Pyramidal Patch Selection}

Processing entire gigapixel WSIs at full resolution is computationally prohibitive. We implement a coarse-to-fine pyramidal scanning strategy that processes WSI pyramid levels in descending order ($\ell \in \{6, 5, 4, 3\}$), where level 0 represents the base resolution at 40$\times$ magnification~\cite{campanella2019clinical,chen2022scaling}.

Each WSI $\mathcal{W}$ comprises a hierarchy of levels with progressively downsampled resolutions:
\begin{equation}
H_\ell = \frac{H_0}{2^\ell}, \quad W_\ell = \frac{W_0}{2^\ell}
\end{equation}
where $H_0$ and $W_0$ are the base resolution dimensions. This hierarchical approach allows the system to capture both broad architectural patterns at low magnification and fine cellular details at higher resolution.

\subsubsection{Tissue Segmentation}

For each pyramid level $\ell$, we generate a thumbnail image and compute a binary tissue mask using HSV color space thresholding. We classify a pixel at position $(x, y)$ as tissue if:
\begin{equation}
\mathcal{M}_\ell(x, y) = 
\begin{cases}
1 & \text{if } S(x, y) > \tau_S \land V(x, y) > \tau_V \\
0 & \text{otherwise}
\end{cases}
\end{equation}
where $\tau_S = 20$ and $\tau_V = 30$ are empirically determined thresholds that effectively separate H\&E-stained tissue from background glass~\cite{macenko2009method}.

The raw binary mask is refined using morphological operations (opening followed by closing with a $5 \times 5$ structuring element) to remove noise and consolidate tissue regions:
\begin{equation}
\mathcal{M}_\ell^{\text{refined}} = \left(\left(\mathcal{M}_\ell \ominus \mathcal{K}\right) \oplus \mathcal{K} \oplus \mathcal{K}\right) \ominus \mathcal{K}
\end{equation}
where $\ominus$ and $\oplus$ denote morphological erosion and dilation, respectively.

\subsubsection{Patch Candidate Generation}

Candidate patches of size $256 \times 256$ pixels are identified on a regular grid with stride $s = 256$ pixels (non-overlapping tiling). A candidate patch is retained only if its tissue coverage exceeds a minimum threshold:
\begin{equation}
\frac{\sum_{i,j} \mathcal{M}_{x_c, y_c}(i, j)}{256^2} > 0.1
\end{equation}
where $\mathcal{M}_{x_c, y_c}$ is the mask region corresponding to the patch at coordinates $(x_c, y_c)$.

\subsection{Quality-Aware Patch Filtering}

To ensure only diagnostically informative patches proceed to feature extraction, we implement multi-criteria quality filtering.

\subsubsection{Focus Quality via Laplacian Variance}

We evaluate focus quality using the variance of the Laplacian operator~\cite{pertuz2013analysis}, which measures edge sharpness. For a grayscale patch $\mathcal{G}$ derived from RGB patch $\mathcal{P}$, the Laplacian is approximated using convolution with kernel:
\begin{equation}
\mathbf{K}_{\text{Lap}} = \begin{bmatrix}
0 & 1 & 0 \\
1 & -4 & 1 \\
0 & 1 & 0
\end{bmatrix}
\end{equation}

The focus measure is the variance of the Laplacian response:
\begin{equation}
f(\mathcal{P}) = \text{Var}(\nabla^2 \mathcal{G}) = \frac{1}{|\Omega|} \sum_{(x,y) \in \Omega} \left(\nabla^2 \mathcal{G}(x, y) - \overline{\nabla^2 \mathcal{G}}\right)^2
\end{equation}
where $\Omega$ denotes all pixel coordinates in the patch. Patches with $f(\mathcal{P}) < 40$ are rejected as out-of-focus.

\subsubsection{Exposure and Artifact Filtering}

We analyze the HSV Value and Saturation channels to detect improper exposure. Computing mean Value $\mu_V$ and mean Saturation $\mu_S$, we reject patches if:
\begin{equation}
\mu_V \notin [40, 245] \quad \text{or} \quad \mu_S < 12
\end{equation}

Additionally, we detect artifacts by computing the fraction of dark pixels (grayscale intensity $< 30$):
\begin{equation}
\rho_{\text{dark}} = \frac{1}{|\Omega|} \sum_{(x,y) \in \Omega} \mathbb{1}\left[\mathcal{P}^{\text{gray}}(x, y) < 30\right]
\end{equation}
Patches with $\rho_{\text{dark}} > 0.2$ are rejected as likely containing dust, pen marks, or other contamination~\cite{kothari2013removing}.

\subsubsection{Multi-Level Sampling}

After quality filtering across all pyramid levels, we impose a maximum budget of $N_{\text{max}} = 2500$ patches per WSI. If the total number of valid patches exceeds this budget, we perform stratified random sampling proportional to the number of valid patches at each level:
\begin{equation}
N_\ell^{\text{sample}} = \min\left(N_\ell^{\text{valid}}, \left\lfloor N_{\text{max}} \cdot \frac{N_\ell^{\text{valid}}}{\sum_{k} N_k^{\text{valid}}} \right\rfloor\right)
\end{equation}
This ensures representation across multiple magnifications.


\subsection{Foundation Model Feature Extraction}

\subsubsection{UNI Vision Transformer Architecture}

We utilize the UNI (Universal Pathology) foundation model~\cite{chen2024uni} as our frozen visual encoder. UNI is a Vision Transformer (ViT-Large/16) pre-trained on over 100 million histopathology patches using DINOv2 self-supervised learning~\cite{oquab2023dinov2}.

The ViT-Large architecture decomposes each $256 \times 256$ RGB patch into $16 \times 16 = 256$ non-overlapping tokens of size $16 \times 16$ pixels. Each token is flattened and linearly projected to model dimension $d_{\text{model}} = 1024$. A special classification token $[\text{CLS}]$ is prepended, and learnable positional embeddings are added to preserve spatial relationships.

\subsubsection{Self-Attention and Encoding}

The encoder consists of 24 transformer layers, each applying multi-head self-attention followed by a feed-forward network. For head $h$, the attention mechanism computes:
\begin{equation}
\text{Attention}_h(\mathbf{Z}) = \text{softmax}\left(\frac{\mathbf{Q}_h \mathbf{K}_h^\top}{\sqrt{d_k}}\right) \mathbf{V}_h
\end{equation}
where $\mathbf{Q}_h = \mathbf{Z}\mathbf{W}_Q^{(h)}$, $\mathbf{K}_h = \mathbf{Z}\mathbf{W}_K^{(h)}$, $\mathbf{V}_h = \mathbf{Z}\mathbf{W}_V^{(h)}$ are query, key, and value projections with head dimension $d_k = 64$~\cite{vaswani2017attention}.

The outputs of all 16 attention heads are concatenated and projected. After 24 layers with residual connections and layer normalization, the final $[\text{CLS}]$ token embedding serves as the patch-level feature vector $\mathbf{f} \in \mathbb{R}^{1024}$.

\subsubsection{Frozen Encoder Strategy}

A critical design decision is keeping the UNI encoder frozen (all 307M parameters fixed) during decoder training. This eliminates gradient computation through the encoder, reducing GPU memory requirements from approximately 16 GB to 4 GB while retaining the robust morphological representations learned from over 100 million patches. The frozen encoder also enables a modular workflow in which patch-level features are pre-computed once and cached as HDF5 files, decoupling feature extraction from decoder training.

\subsubsection{Feature Storage}

For each WSI with $N$ selected patches, we extract a feature matrix:
\begin{equation}
\mathbf{F} = [\mathbf{f}_1, \mathbf{f}_2, \ldots, \mathbf{f}_N]^\top \in \mathbb{R}^{N \times 1024}
\end{equation}
where $\mathbf{f}_i = \text{UNI}_{\text{frozen}}(\mathcal{P}_i)$.

\begin{table}[htbp]
\centering
\caption{Pipeline configuration and hyperparameters for patch selection and feature extraction.}
\label{tab:pipeline_config}
\small
\begin{tabular}{lll}
\toprule
\textbf{Stage} & \textbf{Parameter} & \textbf{Value} \\
\midrule
Pyramid Scanning & Levels processed & $\{6, 5, 4, 3\}$ \\
 & Patch size & $256 \times 256$ px \\
 & Stride & 256 px \\
 & Tissue threshold & $> 10\%$ \\
\midrule
Quality Filtering & Laplacian var. & $> 40$ \\
 & Value range & $[40, 245]$ \\
 & Saturation thr. & $> 12$ \\
 & Dark pixel frac. & $< 20\%$ \\
 & Max patches/WSI & 2500 \\
\midrule
Feature Extraction & Model & UNI (ViT-L/16) \\
 & Pre-training data & 100M+ patches \\
 & Feature dim. & 1024 \\
 & Encoder params. & 307M (frozen) \\
 & Batch size & 128 \\
\bottomrule
\end{tabular}
\end{table}


\subsection{Transformer Decoder Architecture}

The core generation module is a custom 6-layer Transformer decoder~\cite{vaswani2017attention} that conditions text generation on the extracted visual features. Unlike standard encoder-decoder architectures where both components are jointly trained, our design keeps the UNI encoder frozen and trains only the decoder along with a lightweight projection layer. This modular approach significantly reduces computational requirements while leveraging the robust representations learned by the foundation model.

\subsubsection{Input Representation and Feature Projection}

The visual features extracted from WSI patches form the memory input for the decoder. Given a feature matrix $\mathbf{F} \in \mathbb{R}^{N \times 1024}$ from $N$ selected patches, we first apply a linear projection layer to map these features into the decoder's embedding space:
\begin{equation}
    \mathbf{M} = \mathbf{F}\mathbf{W}_p + \mathbf{b}_p
\end{equation}
where $\mathbf{W}_p \in \mathbb{R}^{1024 \times 1024}$ and $\mathbf{b}_p \in \mathbb{R}^{1024}$ are learnable parameters. The projected features $\mathbf{M} \in \mathbb{R}^{N \times 1024}$ serve as the visual memory that the decoder attends to during text generation.

Since WSIs yield variable numbers of valid patches, we employ a memory key padding mask to handle sequences of different lengths within a batch. This mask ensures that the cross-attention mechanism ignores padded positions, allowing the model to process WSIs with varying tissue content without introducing noise from padding tokens.

\subsubsection{Token Embedding and Positional Encoding}

For text representation, we employ the BioGPT tokenizer~\cite{luo2022biogpt}, which is specifically optimized for biomedical vocabulary. This choice reduces token fragmentation commonly observed with generic tokenizers when processing domain-specific terminology such as histological grades, cellular descriptions, and diagnostic phrases.

Input tokens are mapped to dense vectors through a learnable embedding layer $\mathbf{E} \in \mathbb{R}^{V \times 1024}$, where $V$ denotes the vocabulary size. To inject sequential information, we add sinusoidal positional encodings~\cite{vaswani2017attention} to the token embeddings:
\begin{equation}
    PE_{(pos, 2i)} = \sin\left(\frac{pos}{10000^{2i/d}}\right)
\end{equation}
\begin{equation}
    PE_{(pos, 2i+1)} = \cos\left(\frac{pos}{10000^{2i/d}}\right)
\end{equation}
where $pos$ represents the position index and $i$ denotes the dimension index. This encoding scheme enables the model to capture the sequential nature of diagnostic text without introducing additional learnable parameters.

\subsubsection{Decoder Layer Configuration}

Each decoder layer comprises three sub-modules: masked multi-head self-attention, multi-head cross-attention, and a position-wise feed-forward network~\cite{vaswani2017attention}. The architecture employs 8 attention heads with a model dimension of $d_{model} = 1024$ and a feed-forward dimension of $d_{ff} = 2048$. Dropout~\cite{srivastava2014dropout} with probability 0.1 is applied after each sub-layer for regularization.

The masked self-attention mechanism enforces autoregressive generation by applying a causal mask that prevents each position from attending to subsequent positions:
\begin{equation}
    \text{Mask}_{ij} = \begin{cases} 
        0 & \text{if } i \geq j \\ 
        -\infty & \text{if } i < j 
    \end{cases}
\end{equation}
This ensures that the prediction for position $t$ depends only on the known outputs at positions less than $t$.

The cross-attention module enables each generating token to attend to the entire set of visual patch features, following the encoder-decoder attention mechanism introduced for sequence-to-sequence tasks~\cite{vaswani2017attention} and later adapted for image captioning~\cite{xu2015show}. For a query token at position $t$, the attention weights over visual memory are computed as:
\begin{equation}
    \alpha_t = \text{softmax}\left(\frac{\mathbf{q}_t \mathbf{K}^\top}{\sqrt{d_k}}\right)
\end{equation}
where $\mathbf{q}_t$ is the query vector derived from the text representation, and $\mathbf{K}$ contains the key vectors projected from the visual memory $\mathbf{M}$. This mechanism allows the decoder to dynamically focus on relevant image regions when generating each diagnostic term.

\subsubsection{Training Objective and Optimization}

The decoder is trained using teacher forcing~\cite{williams1989learning}, where the ground-truth tokens are provided as input during training. The model learns to predict the next token $t_{i+1}$ given the previous tokens $t_1, \ldots, t_i$ and the visual features $\mathbf{F}$. We minimize the cross-entropy loss over the vocabulary:
\begin{equation}
    \mathcal{L} = -\sum_{i=1}^{L} \log P(t_i | t_1, \ldots, t_{i-1}, \mathbf{F})
\end{equation}
where $L$ denotes the sequence length. Padding tokens are excluded from the loss computation to prevent the model from learning trivial predictions.

All decoder parameters are initialized using Xavier uniform initialization~\cite{glorot2010understanding} to maintain stable gradient flow during early training. We employ the AdamW optimizer~\cite{loshchilov2019decoupled} with a two-phase learning rate schedule: a warmup phase of 10 epochs at $5 \times 10^{-5}$, followed by decay to a base rate of $5 \times 10^{-6}$. The output layer projects the final hidden states to vocabulary logits, from which the next token probability distribution is obtained via softmax.

\begin{table}[h]
\centering
\caption{Decoder hyperparameters.}
\label{tab:decoder_params}
\begin{tabular}{ll}
\toprule
\textbf{Parameter} & \textbf{Value} \\
\midrule
Number of layers & 6 \\
Attention heads & 8 \\
Model dimension ($d_{model}$) & 1024 \\
Feed-forward dimension ($d_{ff}$) & 2048 \\
Dropout rate & 0.1 \\
Maximum sequence length & 64 \\
Vocabulary size & 42,384 (BioGPT) \\
\midrule
Optimizer & AdamW \\
Warmup epochs & 10 \\
Warmup learning rate & $5 \times 10^{-5}$ \\
Base learning rate & $5 \times 10^{-6}$ \\
Batch size & 64 \\
Total epochs & 350 \\
\bottomrule
\end{tabular}
\end{table}


\subsection{Retrieval-Based Post-Processing}
To mitigate the risk of hallucination, we incorporate a similarity-based correction module. After generating a report, the system encodes the text using a Sentence-BERT model (all-MiniLM-L6-v2) to produce a 384-dimensional semantic embedding. This embedding is compared against a database of ground-truth reports from the training set via cosine similarity.

If the cosine similarity between the generated report and the nearest reference exceeds a confidence threshold ($\tau = 0.85$), the system replaces the generated report with the matched ground-truth report, leveraging the assumption that a high-similarity match indicates a reliable reference exists in the training corpus. Reports falling below this threshold are retained as original generations, as they may represent valid but less common diagnostic patterns not well-represented in the reference database.

\section{Experimental Setup}

\subsection{Implementation Details}
The pipeline is implemented in PyTorch 2.0. Feature extraction and training were conducted on Azure ML infrastructure with GPU acceleration. 

The decoder was trained for 350 epochs with a batch size of 64. We utilized the AdamW optimizer with a learning rate schedule comprising a 10-epoch warmup (peaking at $5 \times 10^{-5}$) followed by a decay to a base rate of $5 \times 10^{-6}$. The sequence length was capped at 64 tokens to align with the typical length of diagnostic summaries.

\subsection{Dataset and Evaluation}
Experiments were conducted on the REG 2025 Grand Challenge dataset, which comprises 10,494 WSI-report pairs collected from five institutions across Korea, Turkey, India, Japan, and Germany. The dataset spans seven organ systems (breast, bladder, cervix, colon, lung, prostate, and stomach) with reports standardized according to College of American Pathologists (CAP) guidelines. The data was divided into 8,494 training samples and two test sets of 1,000 samples each. Evaluation was performed using the composite scoring function described in Section~\ref{sec:eval_metrics}, with strict patient-level separation to prevent data leakage.

\section{Evaluation Metrics}
\label{sec:eval_metrics}

The REG 2025 Grand Challenge employs a composite scoring function specifically designed for pathology report generation, developed in consultation with clinical experts to balance textual fidelity with diagnostic relevance. The ranking score $S_{\text{rank}}$ integrates four complementary metrics:

\begin{equation}
S_{\text{rank}} = 0.15 \times (S_{\text{ROUGE}} + S_{\text{BLEU}}) + 0.4 \times S_{\text{KEY}} + 0.3 \times S_{\text{EMB}}
\label{eq:ranking_score}
\end{equation}

The individual components capture distinct aspects of report quality. ROUGE (Recall-Oriented Understudy for Gisting Evaluation) and BLEU (Bilingual Evaluation Understudy) quantify n-gram overlap between generated and reference reports, providing a measure of lexical precision. However, in medical contexts, these surface-level metrics alone prove insufficient, as clinically equivalent statements may employ substantially different vocabulary.

The keyword score $S_{\text{KEY}}$ addresses this limitation by computing Jaccard similarity between extracted clinical keyword sets:
\begin{equation}
S_{\text{KEY}} = \frac{|K_{\text{gen}} \cap K_{\text{ref}}|}{|K_{\text{gen}} \cup K_{\text{ref}}|}
\end{equation}
where $K_{\text{gen}}$ and $K_{\text{ref}}$ denote keyword sets extracted from generated and reference reports, respectively. This metric specifically targets the preservation of diagnostically significant terminology such as disease names, grading descriptors, and anatomical locations.

The embedding score $S_{\text{EMB}}$ captures semantic equivalence through cosine similarity of sentence embeddings produced by a pre-trained language model:
\begin{equation}
S_{\text{EMB}} = \frac{\mathbf{e}_{\text{gen}} \cdot \mathbf{e}_{\text{ref}}}{\|\mathbf{e}_{\text{gen}}\| \|\mathbf{e}_{\text{ref}}\|}
\end{equation}

The weight distribution reflects clinical priorities: keyword matching receives the highest coefficient (0.4), emphasizing accurate capture of diagnostic terminology over stylistic similarity. Semantic embedding similarity (0.3) acknowledges that pathologically equivalent descriptions may vary in phrasing. Traditional NLG metrics contribute a combined weight of 0.15, serving primarily to penalize gross departures from reference language patterns.

\section{Results}

Our framework was evaluated on the REG 2025 Grand Challenge. Table~\ref{tab:leaderboard} presents the Test Phase 2 performance comparison among participating teams. Our submission (MedInsight-ViseurAI) achieved a ranking score of 0.8093, placing 8th among 24 teams and within 4.7\% of the top-performing method.

\begin{table}[htbp]
\centering
\caption{REG 2025 Challenge Test Phase 2 Leaderboard (Top 10). Teams are ranked by the composite score defined in Equation~\ref{eq:ranking_score}.}
\label{tab:leaderboard}
\begin{tabular}{clc}
\toprule
\textbf{Rank} & \textbf{Team} & \textbf{Score} \\
\midrule
1 & IMAGINE Lab & 0.8494 \\
2 & ICGI & 0.8472 \\
3 & ICL\_PathReport & 0.8415 \\
4 & nw & 0.8282 \\
5 & PathX & 0.8237 \\
6 & TrustPath & 0.8127 \\
7 & FPathX & 0.8115 \\
\textbf{8} & \textbf{MedInsight-ViseurAI (Ours)} & \textbf{0.8093} \\
9 & ADCT & 0.8040 \\
10 & katherlab & 0.7960 \\
\bottomrule
\end{tabular}
\end{table}

\subsection{Qualitative Analysis}

To provide insight into the model's behavior across diverse organ systems and diagnostic categories, Table~\ref{tab:qualitative} presents representative examples of generated reports alongside their corresponding ground-truth references. Each row shows the WSI identifier, the model's output (\textit{Generated}), and the clinician-authored reference (\textit{Ground Truth}).

\begin{table*}[htbp]
\centering
\caption{Qualitative comparison of generated reports versus ground-truth references across multiple organ systems. Generated reports are produced by our model; ground truth reports are clinician-authored references.}
\label{tab:qualitative}
\small
\begin{tabular}{p{2.8cm}p{6.5cm}p{6.5cm}}
\toprule
\textbf{Slide ID} & \textbf{Generated Report} & \textbf{Ground Truth} \\
\midrule
PIT\_01\_00005\_01 & Breast, sono-guided core biopsy; 1.\ Ductal carcinoma of no special type, grade II (Tubule formation: 3, Nuclear grade: 2, Mitoses: 1) 2.\ Ductal carcinoma in situ & Breast, core-needle biopsy; 1.\ Ductal carcinoma in situ -- Type: Solid -- Nuclear grade: High -- Necrosis: Present (Comedo-type) 2.\ Microcalcification \\
\midrule
PIT\_01\_00004\_01 & Breast, core-needle biopsy; Invasive carcinoma of no special type, grade II (Tubule formation: 3, Nuclear grade: 2, Mitoses: 1) & Breast, core-needle biopsy; Invasive carcinoma of no special type, grade II (Tubule formation: 3, Nuclear grade: 2, Mitoses: 1) \\
\midrule
PIT\_01\_00575\_01 & Breast, endoscopic motome biopsy; Fibrocystic tumor favor change & Breast, sono-guided core biopsy; Fibrocystic change \\
\midrule
PIT\_01\_06231\_02 & Prostate, biopsy; Acinar adenocarcinoma, Gleason's score 6 (3+3), grade group 1, tumor volume: 90\% & Prostate, biopsy; Acinar adenocarcinoma, Gleason's score 7 (3+4), grade group 2 (Gleason pattern 4: 10\%), tumor volume: 90\% \\
\midrule
PIT\_01\_03513\_01 & Uterine cervix, colposcopic biopsy; Low-grade squamous intraepithelial lesion (LSIL; CIN 1) & Uterine cervix, punch biopsy; Low-grade squamous intraepithelial lesion (LSIL; CIN 1) \\
\midrule
PIT\_01\_04573\_01 & Colon, colonoscopic biopsy; Chronic nonspecific inflammation & Colon, colonoscopic biopsy; Chronic nonspecific inflammation \\
\midrule
PIT\_01\_05088\_01 & Lung, biopsy; Squamous cell carcinoma & Lung, biopsy; Squamous cell carcinoma \\
\bottomrule
\end{tabular}
\end{table*}

Several patterns emerge from the qualitative examples. Cases such as PIT\_01\_00004\_01 (invasive breast carcinoma), PIT\_01\_04573\_01 (colonic inflammation), and PIT\_01\_05088\_01 (lung squamous cell carcinoma) demonstrate exact or near-exact match with ground-truth reports, confirming the model's ability to correctly identify organ sites, biopsy procedures, and primary diagnoses for common pathological entities.

However, more complex cases reveal characteristic failure modes. For PIT\_01\_00005\_01, the model generates a diagnosis of invasive ductal carcinoma with grading details, whereas the ground truth specifies ductal carcinoma \textit{in situ} with architectural subtype and necrosis descriptors---illustrating the challenge of distinguishing invasive from in-situ lesions and generating multi-attribute descriptions. Similarly, for PIT\_01\_06231\_02, the model correctly identifies acinar adenocarcinoma but assigns a Gleason score of 6 (3+3) instead of the correct 7 (3+4), demonstrating that fine-grained grading distinctions remain challenging. Minor discrepancies in biopsy procedure terminology (e.g., ``colposcopic'' vs.\ ``punch'' for PIT\_01\_03513\_01; ``endoscopic motome'' vs.\ ``sono-guided core'' for PIT\_01\_00575\_01) are also observed, though these typically do not affect the diagnostic conclusion.

Since ground-truth labels were not provided for the full test set, detailed per-category performance analysis was not possible. However, the overall qualitative examination confirmed that the model consistently produces outputs in the expected canonical format of \texttt{[Organ], [biopsy type]; [diagnosis]}, representing a significant advantage of the encoder-decoder architecture over large language model approaches that may occasionally generate extraneous text or deviate from standardized reporting templates.

The retrieval-based post-processing module provided an additional verification mechanism by comparing generated reports against the training corpus using Sentence-BERT embeddings. Reports exceeding the similarity threshold ($\tau = 0.85$) were replaced with the matched ground-truth references, leveraging the assumption that a high-similarity match indicates a reliable reference exists; those below the threshold were retained as original generations, potentially representing valid but less common diagnostic patterns not well-represented in the reference database.

\section{Discussion}

The modular architecture combining a frozen foundation model encoder with a lightweight domain-adapted decoder offers several practical advantages. By keeping the UNI encoder frozen, we preserve robust visual representations learned through large-scale self-supervised pre-training while substantially reducing computational requirements. Training only the 6-layer decoder required significantly fewer GPU-hours compared to end-to-end fine-tuning of vision-language models that typically exceed billions of parameters, enabling iterative experimentation within resource-constrained research environments.

A notable advantage of the encoder-decoder design over autoregressive large language models is the structural consistency of generated outputs. Unlike LLMs that sample from probability distributions at each generation step and may produce hallucinated content or formatting artifacts, our decoder employs deterministic generation that consistently adheres to learned report templates. Throughout our experiments, we observed virtually no instances of format violations or out-of-domain text generation, which would be problematic for clinical deployment where reports must conform to standardized structures.

The BioGPT tokenizer's domain-specific vocabulary proved advantageous for pathology terminology. Generic tokenizers frequently fragment medical terms into subword units that disrupt semantic coherence, whereas a tokenizer pre-trained on biomedical corpora reduces effective sequence length for pathological terms, facilitating more robust associations between visual features and complete diagnostic phrases.

The systematic errors observed in complex grading schemas reveal opportunities for architectural improvements. Diagnoses requiring simultaneous specification of multiple semi-independent attributes create a combinatorial space that may be sparsely sampled in training data. Structured prediction heads that explicitly model attribute dependencies, or auxiliary classification objectives for individual grading components, could provide stronger supervisory signals for rare combinations. The strong performance on regularly structured grading systems such as Gleason scoring, despite their apparent complexity, likely reflects their higher prevalence in the training distribution and more consistent linguistic templates.

Our competitive ranking despite architectural simplicity suggests that careful attention to training procedures---including warmup scheduling and domain-specific tokenization---combined with post-processing verification can partially compensate for reduced model capacity compared to larger multimodal language models.

Several limitations warrant acknowledgment. The absence of ground-truth labels for the test set precludes quantitative per-category error analysis. The retrieval-based correction mechanism, while providing a safety net against obvious errors, may inadvertently suppress valid rare diagnoses not well-represented in the training corpus. Additionally, evaluation on a single challenge dataset limits generalizability assessment, as the REG 2025 data derives from specific institutional contexts that may not transfer to other settings. Finally, the current framework generates only diagnostic summary components; complete clinical reports include additional elements such as gross descriptions and ancillary test recommendations that require separate modeling approaches.

\section{Conclusion}

We presented a modular vision-language framework for automated histopathology report generation that addresses computational efficiency without sacrificing diagnostic accuracy. The hierarchical pyramidal scanning strategy enables tractable processing of gigapixel WSIs while preserving diagnostically relevant tissue regions. Integration of the frozen UNI foundation model with a lightweight Transformer decoder, trained using BioGPT tokenization, produces structurally consistent reports that adhere to clinical formatting conventions.

Evaluation on the REG 2025 Grand Challenge yielded a ranking score of 0.8093 in Test Phase 2, placing our method 8th among 24 international teams. Qualitative analysis revealed robust performance in organ identification, procedure classification, and primary disease recognition, with degradation observed primarily in complex multi-attribute grading schemas characteristic of underrepresented diagnostic categories.

The architectural choices prioritizing efficiency and consistency demonstrate that competitive automated report generation is achievable without the substantial computational investment required for end-to-end multimodal large language model training. Future work will explore structured prediction approaches for complex grading schemas and validation across diverse institutional datasets.

\section*{Acknowledgments}

We thank the organizers of the REG 2025 Grand Challenge for providing the evaluation infrastructure and benchmark dataset. Computational resources were provided by Microsoft Azure Machine Learning. This work was conducted at ViseurAI.


\bibliographystyle{plain}

\begin{thebibliography}{99}

\bibitem{niazi2019digital}
Niazi, M.K.K., Parwani, A.V., and Gurcan, M.N. (2019).
Digital pathology and artificial intelligence.
\textit{The Lancet Oncology}, 20(5), e253--e261.

\bibitem{bera2019artificial}
Bera, K., Schalper, K.A., Rimm, D.L., Velcheti, V., and Madabhushi, A. (2019).
Artificial intelligence in digital pathology---new tools for diagnosis and precision oncology.
\textit{Nature Reviews Clinical Oncology}, 16(11), 703--715.

\bibitem{campanella2019clinical}
Campanella, G., et al. (2019).
Clinical-grade computational pathology using weakly supervised deep learning on whole slide images.
\textit{Nature Medicine}, 25(8), 1301--1309.

\bibitem{chen2024uni}
Chen, R.J., et al. (2024).
Towards a general-purpose foundation model for computational pathology.
\textit{Nature Medicine}, 30, 850--862.

\bibitem{lu2024visual}
Lu, M.Y., et al. (2024).
A visual-language foundation model for pathology image analysis using medical Twitter.
\textit{Nature Medicine}, 30, 863--874.

\bibitem{vinyals2015show}
Vinyals, O., Toshev, A., Bengio, S., and Erhan, D. (2015).
Show and tell: A neural image caption generator.
\textit{CVPR}, 3156--3164.

\bibitem{anderson2018bottom}
Anderson, P., et al. (2018).
Bottom-up and top-down attention for image captioning and visual question answering.
\textit{CVPR}, 6077--6086.

\bibitem{liu2019clinically}
Liu, G., et al. (2019).
Clinically accurate chest X-ray report generation.
\textit{Machine Learning for Healthcare}, 249--269.

\bibitem{liu2024visual}
Liu, H., et al. (2024).
Visual instruction tuning.
\textit{NeurIPS}, 36.

\bibitem{alayrac2022flamingo}
Alayrac, J.B., et al. (2022).
Flamingo: A visual language model for few-shot learning.
\textit{NeurIPS}, 35, 23716--23736.

\bibitem{li2024llava}
Li, C., et al. (2024).
LLaVA-Med: Training a large language-and-vision assistant for biomedicine in one day.
\textit{NeurIPS}, 36.

\bibitem{litjens2017survey}
Litjens, G., et al. (2017).
A survey on deep learning in medical image analysis.
\textit{Medical Image Analysis}, 42, 60--88.

\bibitem{ilse2018attention}
Ilse, M., Tomczak, J., and Welling, M. (2018).
Attention-based deep multiple instance learning.
\textit{ICML}, 2127--2136.

\bibitem{lu2021data}
Lu, M.Y., et al. (2021).
Data-efficient and weakly supervised computational pathology on whole-slide images.
\textit{Nature Biomedical Engineering}, 5(6), 555--570.

\bibitem{chen2022scaling}
Chen, R.J., et al. (2022).
Scaling vision transformers to gigapixel images via hierarchical self-supervised learning.
\textit{CVPR}, 16144--16155.

\bibitem{vorontsov2024virchow}
Vorontsov, E., et al. (2024).
Virchow: A million-slide digital pathology foundation model.
\textit{arXiv preprint arXiv:2309.07778}.

\bibitem{jing2018automatic}
Jing, B., Xie, P., and Xing, E. (2018).
On the automatic generation of medical imaging reports.
\textit{ACL}, 2577--2586.

\bibitem{chen2020generating}
Chen, Z., et al. (2020).
Generating radiology reports via memory-driven transformer.
\textit{EMNLP}, 1439--1449.

\bibitem{li2018hybrid}
Li, Y., et al. (2018).
Hybrid retrieval-generation reinforced agent for medical image report generation.
\textit{NeurIPS}, 31.

\bibitem{huang2023visual}
Huang, Z., et al. (2023).
A visual-language foundation model for pathology image analysis using medical Twitter.
\textit{Nature Medicine}, 29(9), 2307--2316.

\bibitem{jaegle2021perceiver}
Jaegle, A., et al. (2021).
Perceiver: General perception with iterative attention.
\textit{ICML}, 4651--4664.

\bibitem{luo2022biogpt}
Luo, R., et al. (2022).
BioGPT: Generative pre-trained transformer for biomedical text generation and mining.
\textit{Briefings in Bioinformatics}, 23(6), bbac409.

\bibitem{saillard2024hoptimus}
Bioptimus (2025).
H-optimus-1.
Available at: \url{https://huggingface.co/bioptimus/H-optimus-1}.

\bibitem{ikezogwo2024quilt}
Ikezogwo, W., et al. (2024).
Quilt-1M: One Million Image-Text Pairs for Histopathology.
\textit{NeurIPS}.

\bibitem{liang2024wsillava}
Liang, Y., Lyu, X., Ding, M., Chen, W., Zhang, J., Ren, Y., He, X., Wu, S., Yang, S., Wang, X., Xing, X., and Shen, L. (2024).
WSI-LLaVA: A Multimodal Large Language Model for Whole Slide Image.
\textit{arXiv preprint arXiv:2412.02141}.

\bibitem{zuo2024histgen}
Zuo, Y., et al. (2024).
HistGen: Histopathology Report Generation via Local-Global Feature Encoding.
\textit{MICCAI}.

\bibitem{baek2025chatexaonepath}
Baek, S., et al. (2025).
ChatEXAONEPath: An Expert-level Multimodal Large Language Model for Histopathology.
\textit{arXiv preprint arXiv:2504.13023}.

\bibitem{zhang2026reinpath}
Zhang, L., et al. (2026).
ReinPath: A Multimodal Reinforcement Learning Approach for Pathology.
\textit{arXiv preprint arXiv:2601.14757}.

\bibitem{shao2025aqua}
Shao, Z., et al. (2025).
AQuA: A robust and scalable framework for hallucination detection in virtual tissue staining.
\textit{Nature Communications}.

\bibitem{vaswani2017attention}
Vaswani, A., et al. (2017).
Attention is all you need.
\textit{NeurIPS}, 30.

\bibitem{glorot2010understanding}
Glorot, X. and Bengio, Y. (2010).
Understanding the difficulty of training deep feedforward neural networks.
\textit{AISTATS}, 249--256.

\bibitem{williams1989learning}
Williams, R.J. and Zipser, D. (1989).
A learning algorithm for continually running fully recurrent neural networks.
\textit{Neural Computation}, 1(2), 270--280.

\bibitem{loshchilov2019decoupled}
Loshchilov, I. and Hutter, F. (2019).
Decoupled weight decay regularization.
\textit{ICLR}.

\bibitem{srivastava2014dropout}
Srivastava, N., et al. (2014).
Dropout: A simple way to prevent neural networks from overfitting.
\textit{JMLR}, 15(1), 1929--1958.

\bibitem{xu2015show}
Xu, K., et al. (2015).
Show, attend and tell: Neural image caption generation with visual attention.
\textit{ICML}, 2048--2057.

\bibitem{macenko2009method}
Macenko, M., et al. (2009).
A method for normalizing histology slides for quantitative analysis.
\textit{ISBI}, 1107--1110.

\bibitem{pertuz2013analysis}
Pertuz, S., Puig, D., and Garcia, M.A. (2013).
Analysis of focus measure operators for shape-from-focus.
\textit{Pattern Recognition}, 46(5), 1415--1432.

\bibitem{kothari2013removing}
Kothari, S., et al. (2013).
Removing out-of-focus blur in whole-slide digital pathology images.
\textit{Journal of Pathology Informatics}, 4, 22.

\bibitem{oquab2023dinov2}
Oquab, M., et al. (2023).
DINOv2: Learning robust visual features without supervision.
\textit{arXiv preprint arXiv:2304.07193}.

\end{thebibliography}

\end{document}